# The Effects of Early Collisional Evolution on Amorphous Water Ice Bodies

Jordan K. Steckloff[1,2], Gal Sarid[3], Brandon C. Johnson[4,5]

[1]Planetary Science Institute, Tucson, AZ
[2]University of Texas at Austin, Austin, TX
[3]SETI Institute, Mountain View, CA
[4]Department of Earth, Atmospheric, and Planetary Sciences, Purdue University, West Lafayette, IN
[5]Department of Physics and Astronomy, Purdue University, West Lafayette, IN

**ABSTRACT**

Conditions in the outer protoplanetary disk during Solar System formation were thought to be favorable for the formation of amorphous water ice (AWI), a glassy phase of water ice. However, subsequent collisional processing could have shock crystallized any AWI present. Here we use the iSALE shock physics hydrocode to simulate impacts between large icy bodies at impact velocities relevant to these collisional environments, and then feed these results into a custom-built AWI crystallization script, to compute how much AWI crystallizes/survives these impact events. We find that impact speeds between icy bodies post-planet migration (i.e., between trans-Neptunian Objects or TNOs) are too slow to crystallize any meaningful fraction of AWI. During planet migration, however, the amount of AWI that crystallizes is highly stochastic: relatively little AWI crystallizes at lower impact velocities (less than ~2 km/s), yet most AWI present in the bodies (if equal sized) or impactor and impact site (if different sizes) crystallizes at higher impact velocities (greater than ~4 km/s). Given that suspected impact speeds during planet migration were ~2-4 km/s, this suggests that primordial AWI's ability to survive planet migration is highly stochastic. However, if proto-EKB objects and their fragments experienced multiple impact events, nearly all primordial AWI could have crystallized; such a highly collisional proto-EKB during planet migration is consistent with the lack of any unambiguous direct detection of AWI on any icy body. Ultimately, primordial AWI's survival to the present day depends sensitively on the proto-EKB's size-frequency distribution, which is currently poorly understood.

1.  **Introduction**

Amorphous water ice (AWI) is a solid, glassy phase of water ice that forms in the conditions of the solar nebular and proto-planetary disk (*Mastrapa et al. 2013; Ciesla, 2014*). Models suggest that AWI later adsorbed onto silicate grain surface as complete icy grains, or was included into mixed silicate grains and accreted together to form the original population of cometary bodies in the outer solar system (*e.g., Weidenschilling, 2004*), and may survive to the present day. Nevertheless, although two tentative, weak spectroscopic detections of AWI have been reported (*Davies et al. 1997; Kawakita et al. 2004*), primordial AWI has never been conclusively detected on an icy body's surface in the present Solar System (*Lisse et al. 2013*)[1]. The detection of $N_2$, $O_2$, and Ar on comet 67P/Churyumov-Gerasimenko (*e.g., Davidsson et al. 2016*), requires a very cold nucleus interior (<30 K). However, debate is still active in determining whether the solid water phase in small icy bodies is amorphous, crystalline, or a form of clathrate hydrates (*e.g., Meech & Svoren 2004; Marboeuf et al. 2012; Lisse et al. 2013; Choukroun et al. 2013; Mastrapa et al. 2013; Luspay-Kuti et al. 2016; Brugger et al. 2016*).

Pure amorphous water ice can spontaneously and irreversibly transition into crystalline water ice once warmed above ~120 K, releasing a significant amount of energy on the order of ~$10^5$ J/kg (*e.g., Ghormley, 1968; Klinger, 1981*). The release of this enthalpy of crystallization can further drive up the temperature of ice by ~45 K, potentially triggering further crystallization and inducing runaway crystallization (*e.g., Smoluchowski 1981; Prialnik et al. 2004*). Accordingly, much of the outermost ice of a comet that has been heated to ~140 K is expected to be in the crystalline form.

Because of these energetics, the crystallization of AWI is often thought to be the energy source driving cometary outbursts, explosions and dust jet activity (*e.g., Prialnik & Bar-Nun 1992; Meech & Svoren 2004; Prialnik et al. 2004; Sarid et al. 2005; Prialnik et al. 2008b;a; Sekanina 2009; Jewitt 2009; Kossacki & Szutowicz 2013; Bruck Syal et al. 2013; Ishiguro et al. 2014; Mousis et al. 2015; Guilbert-Lepoutre et al. 2015; Agarwal et al. 2017*). Additionally, AWI crystallization has been attributed to the production of common cometary highly-volatile species (e.g., CO and $CO_2$) and noble gasses, due to the ability of AWI to trap such molecules and release it upon crystallization (*e.g., Bar-Nun et al. 1985; 2007; Collings et al. 2003*).

However, AWI-free mechanisms have been proposed to explain all of these cometary behaviors (*e.g., Crifo et al. 2002; Kossacki & Szutowicz 2011; Combi et al. 2012; Grün et al. 2016; Steckloff & Melosh 2016; Steckloff et al. 2016; Steckloff & Samarasinha, 2018*), as well as steep-sided pit formation (*e.g., Britt et al. 2004; Brownlee et al. 2004; Steckloff & Samarasinha, 2018*) and presence of highly-volatile species (*e.g., Gortsas et al. 2011*). Additionally, impurities in AWI strongly reduce the exothermicity of its crystallization, and can even render the crystallization *endothermic* if concentrations of common volatile such as CO reach a few mole percent (*Kouchi & Sirono, 2001*).

---

[1] AWI has been detected on the trailing hemispheres of some Galilean moons, however, this is thought to have formed radiolytically through interactions with trapped particles in Jupiter's magnetic field (*Hansen and McCord, 2004; Ligier et al. 2016*).

Furthermore, thermophysical and collisional evolution may severely affect these ice structures, specifically destroying AWI at the surface and buried at depth. In particular, Nice-style instabilities in the early Solar System (*Tsiganis et al. 2005; Morbidelli et al. 2005; Morbidelli, 2010; Levison et al. 2011; Morgan et al. 2021*) are thought to have lead to significant collisional evolution of the primordial comet population (*Morbidelli and Rickman, 2015*) prior to ejection into the Oort Cloud or dynamically "hot" Trans-Neptunian object population (i.e., the Scattered Disk and Hot Classical Edgeworth-Kuiper Belt objects)[2]. The cratering record on Arrokoth and Charon reveal that the size-frequency distribution of these populations are indeed significantly collisionally evolved (*Singer et al. 2019; Robbins and Singer, 2021; Morbidelli et al. 2021*), consistent with this early collisional evolution during planet migration. Such collisional evolution may cause significant shock-induced crystallization of any primordial abundance of AWI in icy bodies. This begs the question: could amorphous water ice even survive this early collisional environment?

## 2. Methods

To address this question, we use the iSALE hydrodynamics shock physics code to simulate the impact process, and track the thermodynamic conditions of the materials throughout each object during the impact process. We then feed these thermodynamic conditions into an AWI crystallization script, based on published Gibbs Free Energy crystallization models (*Kouchi et al. 1994*). Together, these two components allow us to calculate the amount of AWI that crystallizes as a function of impact parameters (impact speed, initial temperature, object size, and object size ratio).

### 2.1. The iSALE Hydrocode

The iSALE impact shock physics is an arbitrary Lagrangian-Eulerian shock-physics code based on the SALE hydrodynamics code (*Amsden et al. 1980*). iSALE expands upon SALE to include an elastic-plastic constitutive model for impacts into solid bodies, material fragmentation models, multiple materials and their equations of state (*Melosh et al. 1992; Ivanov et al. 1997*), and modified strength models (*Collins et al. 2004*). More recently, the creation of porosity through dilatancy (*Collins 2014*) and porous compaction of materials (*Wünnemann et al. 2006; Collins et al. 2011*) have been incorporated into iSALE, tested against experimental data, and benchmarked against other impact codes (*e.g., Pierazzo et al. 2008*).

The number of high quality equations of state available for use in shock physics codes is limited. There is currently insufficient experimental data on the shock response of amorphous water ice, which precludes the development of an equation of state of amorphous water ice for use in impact hydrocodes such as iSALE. Thus, we must use an equation of state for water ice, which can reasonably approximate the behavior of amorphous water ice during the impact process, due to their similar densities. The Eka-ANEOS equation of state for water (*Turtle & Pierazzo, 2001*) describes the thermodynamic conditions of interest (temperature, pressure, multiple phases) more accurately for our purposes than other available EOSs in iSALE. Although the 5-phase ice EOS of Senft and Stewart (*2008*) based on the experiments of Stewart and Ahrens (*2005*) includes multiple phase transitions, it predicts the same volume of impact melt as the Eka-ANEOS (*Kraus et al. 2011*). Thus, we expect the shock heating of material is not significantly affected by our choice of Eka-ANEOS over the 5-phase ice EOS.

---

[2] This excludes Cold Classical Kuiper Belt Objects, which are thought to have been dynamically unaffected by such collisional evolution during planet migration, as evidenced by the survival of Arrokoth's shape (*McKinnon et al. 2020*)

Although the crystallization of amorphous water ice is an exothermic process (and this approximation assumes that AWI crystallization is energetically inert), its specific enthalpy of crystallization (1.3 kJ/mol) is small compared to the specific kinetic energy of typical impact events (9-900 kJ/mol at 1-10 km/s). Furthermore, this exothermicity disappears at modest impurity concentrations of ~2-3% for common cometary volatiles such as CO, $CO_2$, and $CH_4$ (*Kouchi & Sirono, 2001*). Therefore, neglecting the heat of crystallization in our calculations would likely produce only small errors in the thermodynamic conditions during the impact process, and can be safely neglected. We ignore all phases of water ice other than AWI and ice I, including high pressure phases of ice and the low temperature orthorhombic ice XI. We use the same strength model and associated parameters as Bray et al. *(2014)*.

Furthermore, we assume that these objects initially have zero porosity for simplicity. This assumption seems counter to observations of comet nuclei, which tend to exhibit significant porosity of ~70% (*A'Hearn et al. 2005; Ernst & Schultz, 2007; Pätzold et al. 2016*). However, these nuclei are typically only a few km across, and it is unclear if this porosity is the result of processing such as collisions (*Schwartz et al. 2018*) or rotational disruption (*Safrit et al. 2021*), which could allow these small icy bodies to retain extremely high porosity as they reform. In any case, these results are unlikely to be representative of larger ~100 km objects due to self-gravity, even if such large objects may be able to retain significant porosity. Jupiter Family Comet (JFC) nuclei have bulk strengths on the order of ~1-100 Pa (*Sekanina and Yeomans, 1985; Asphaug and Benz, 1996; Thomas et al. 2015; Steckloff et al. 2015; Hirabayashi et al. 2016; Attree et al. 2018*), which is able to resist the hydrostatic pressures in the interiors of their nuclei, which can reach ~10-100 Pa at their centers. However, the hydrostatic pressures in the interiors of ~100 km objects is on the order of ~1 MPa, which is sufficient to crush out much of the porosity found in these JFCs. Thus, while ~100 km objects can retain significant porosity, it would be expected that their bulk porosities are nevertheless much lower than those found within ~1 km JFCs, consistent with previous laboratory studies (*Yasui and Arakawa, 2009*). This is also consistent with Cassini spacecraft observations of Phoebe, the only ~100 km icy body to be studied in situ, which constrained its density to 1638±33 kg/m$^3$ (*Thomas, 2010*); if Phoebe has a similar bulk composition to Pluto or Triton (objects thought to have formed from Phoebe-like objects), then its bulk porosity is ~15% (*Johnson & Lunine, 2005*). Because porosity should enhance heating (and therefore crystallization), we note that our assumptions should produce conservative results that likely underestimate the full amount of AWI that crystallizes during each impact, as crushing out pore space would lead to enhanced heating and thus greater AWI crystallization. Ultimately, we leave the consideration of pre-impact bulk porosity to our follow-up manuscript, but note that our assumptions should produce conservative results.

We use iSALE's lagrangian tracer particles ("tracers") to record the temperature and pressures over time of the parcel of ice in which each tracer is embedded. As tracer's traverse grid cells with discontinuous properties (e.g., different temperatures), the tracer's thermodynamic profile may record a "jump" in a particular property such as temperature (see figure 2); this behavior is an unavoidable consequence of iSALE's cell based structure. Nevertheless, with sufficiently high grid spacing, the effects of such jumps are small and manageable. However, as grid resolution increases, the runtime of each simulation increases by the cube of the resolution/linear number of grid cells.  Thus, high resolution must be balanced with the limits of the computational resources, to keep total runtimes manageable. We find that using a resolution of 250 grid-cells per object radius gives sufficient resolution while keeping runtimes to a few weeks.

The impact shockwave passes through the target at a rate comparable to the speed of sound. Because the peak shock heating is associated with the passage of this shockwave (*Melosh, 1989*) and the AWI crystallization timescale depends exponentially on temperature, significant errors could be introduced into our calculation from cells whose peak shock temperature corresponds to a crystallization timescale comparable to errors in peak shock duration. Therefore, care must be taken in choosing a recording cadence for the tracers that is fast relative to the passage of the shockwave through the numerical grid cells. In practice, this corresponds to a very narrow range of a few Kelvin; very few grid cells would have a peak temperature in such a narrow range. Thus, our computation of AWI crystallization is insensitive to this data output frequency, so long as they can determine whether or not a cell's peak shock temperature is either above or below this narrow range of peak temperatures. We output data every 0.25 seconds (every 0.1 second for our simulations of a 1km impactors), ensuring that our model records the tracer thermodynamic conditions with sufficient accuracy.

Finally, our model neglects radiative cooling for the initial impactor, target, and ejected materials. This is negligible for the impactor and target, but could become important for ejecta. The longest of our iSALE simulations last 40 minutes; we therefore calculate the maximum particle size for which radiative cooling could be significant. We find that, even if initially at 150 K, all pieces of material larger than ~10 cm would radiatively cool by less than 1 K during a 40 minute simulation. Therefore, radiative cooling is negligible for our simulations.

2.2.    *The Gibbs Free Energy Crystallization Model*
Phase transitions such as amorphous ice crystallization are controlled by the chemical potential ($\mu$) of a species phase, with molecules spontaneously moving into phases that minimize their chemical potential. However, the chemical potential (and therefore the rate of a phase transition) is strongly dependent on both temperature and enthalpy of crystallization. The chemical potential is the molar partial derivative of the Gibbs free energy ($G$) at constant temperature ($T$) and pressure ($p$), given by

$$\mu_i = \left(\frac{\partial G}{\partial N_i}\right)_{T,p}. \tag{1}$$

Although AWI is at a higher chemical potential than crystalline water ice, it is formed under conditions of low pressure and temperature, such that the water molecules are frozen in place faster than they can randomly diffuse into lattice sites to form a crystalline configuration. In other words, the distance that a water molecule may diffuse during deposition is smaller than an ice crystal's unit cell (*Kouchi et al. 1994*). Thus, even though the crystalline phase of water ice is energetically favored over the amorphous phase (lower chemical potential), the molecules are frozen in place before they can find the crystalline structure.

However, thermal energy is unevenly distributed amongst the ice molecules according to the equipartition theorem; a temperature-dependent fraction of the ice molecules will have enough kinetic energy to break their bonds and diffuse into a crystalline structure. Because the chemical potential of crystalline ice is always lower than that of amorphous water ice, ice will remain crystalline once crystalized (*Speedy et al. 1996*). Kouchi et al. (*1994*) used these considerations to derive the crystallization rate of amorphous water ice using the kinetic theory of crystallization (*e.g., Hobbs, 1974*), and benchmarked their equations against experimental data. Their result is an equation for the crystalline fraction ($\theta$) of water ice

$$\theta_{(t)} = 1 - e^{\left(\frac{t}{\tau}\right)^4} \tag{2}$$

$$\tau = \left(\frac{1}{2\pi\alpha}\right)^{\frac{1}{4}} \left(\frac{kT}{\sigma}\right)^{\frac{1}{8}} \frac{\Omega^{\frac{2}{3}}}{D_0} e^{\frac{1}{kT}\left[E_\alpha + \frac{4\pi\sigma^3}{3L^2}\left(\frac{T_m}{T_m-T}\right)^2\right]} \tag{3}$$

where $\tau$ is the crystallization timescale, $\alpha$ is a geometrical factor that depends on the morphology of crystal growth, $\Omega$ is the effective volume of a water molecule, $\sigma = \gamma\Omega^{\frac{2}{3}}$ where $\gamma$ is the interfacial tension (i.e., surface tension), $D_0$ is an empirically derived reference diffusion constant, $E_\alpha$ is the activation energy of self-diffusion, $L$ is the enthalpy (latent heat) of crystallization per molecule at 0 K, and $T_m$ is the freezing point temperature, where solid and liquid phases coexist (*Kouchi et al. 1994*). We plot the temperature dependence of this crystallization timescale in figure 1.

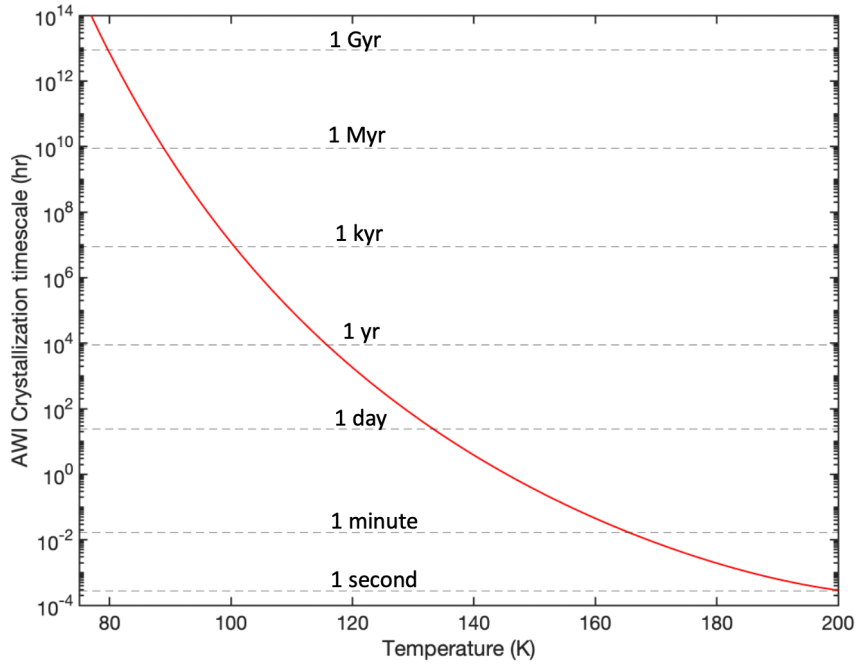

**Figure 1:** *Crystallization timescale of amorphous water ice as a function of temperature.* This is the amorphous water ice (AWI) crystallization timescale from *Kouchi et al. (1994)*, showing that the timescale of amorphous water ice (AWI) crystallization is highly temperature dependent.

This crystallization model assumes that a sample of amorphous water ice is exposed to a constant thermal environment. To adapt this model for the changing thermal environment of the impact process, we created a finite differential form of these equations

$$\theta_{(t+\Delta t)} = \theta_{(t)} + \Delta\theta_{(t)}\left(1 - \theta_{(t)}\right) \quad (4)$$

$$\Delta\theta_{(t)} = 4\left(\frac{\Delta t}{\tau_{(t)}}\right)^4 e^{\left(\frac{\Delta t}{\tau_{(t)}}\right)^4} \quad (5)$$

where $\Delta\theta_{(t)}$ is the fraction of remaining amorphous water ice that crystallizes in a single time step $\Delta t$, which depends on the time-varying crystallization timescale $\tau_{(t)}$. Thus,

$$\theta_{(t)} = \theta_0 + \sum_{t=t_0}^{t} \Delta\theta_{(t-\Delta t)}\left(1 - \theta_{(t-\Delta t)}\right) \quad (6)$$

where $\theta_0$ is the initial crystallized fraction of amorphous water ice.

To calculate the effects of impact processes on the crystallization of amorphous water ice, we feed the thermal history of each iSALE tracer through this crystallization model. This calculates the amount of AWI that crystallizes throughout the process, and is specific to each tracer. We then also aggregate these tracer AWI crystallization fractions, to determine how much of the initial AWI within both impacting objects crystallizes and the degree to which this AWI crystallizes. The degree of crystallization measures the fraction of a parcel of initial AWI that crystallized during the impact process; 0% means that all initial AWI survived the impact and 100% means that all AWI crystallized. In practice, we find that little of the initial AWI takes values in the middle of this range, as nearly all of the initial AWI is either fully crystallized (near 100%) or remains essentially uncrystallized (near 0%).

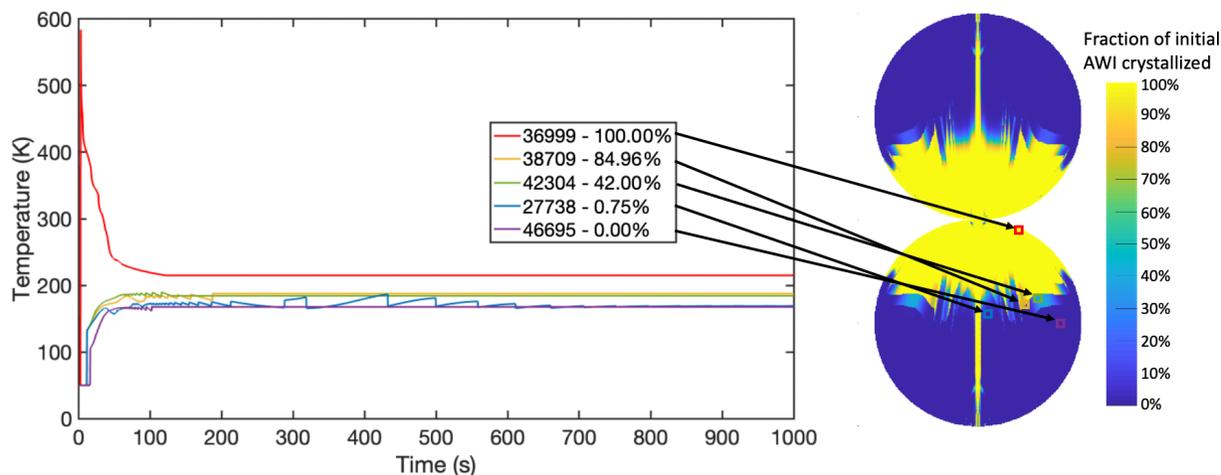

**Figure 2:** *Tracer thermal profiles and locations in colliding bodies.* We plot the thermal profiles for five of the 49212 tracers in our simulation of two 100 km bodies at 50K colliding at 3 km/s. *(Left)* tracers in different parts of the body experience very different thermal profiles. Minor numerical artifacts are visible from the tracer passing through different grid cells, producing a sawtooth pattern in some of the profiles. Tracers are identified by tracer number and the fraction of that tracer's initial AWI that is crystallized. *(Right)* We trace the tracers to their original locations in the bodies at the moment of impact. Colors denote the fraction of each tracer's initial AWI that crystallizes during the impact process.

### 2.3. Simulation Parameters

We conduct a sensitivity study to identify which of the following parameters most affect impact crystallization of amorphous water ice in the early solar system: object size, impact speed, initial temperature, and object size ratio (see Table 1); we then further investigate parameters that sensitively affect the amount of AWI that crystallizes. We conducted simulations between equal-sized objects 25 - 400 km in diameter, as ~100km objects are thought to have dominated the early mass distribution budget of primordial planetesimals (the "Born Big" hypothesis; *Morbidelli et al. 2009*). Here we assume that these objects were initially as a single uniform temperature throughout their interiors. We also assumed that these objects were composed of pure water ice, justifying our use of the Eka-ANEOS equation of state for water (*Turtle & Pierazzo, 2001*). We use a grid resolution of 250 cells per object radius (500 cells per object diameter), resulting in a resolution of 200m for our 100km object. iSALE tracer particles are placed in every other grid cell, resulting in tracers filling the volumes of the colliding objects, spaced 400m apart. To explore the effects of object size on AWI crystallization, we also vary the size of the colliding objects from 25 km to 400 km, scaling the tracer spacing and cell size proportionally to preserve the number of cells and number of tracers.

Impact speeds during collisional evolution of the proto-Kuiper Belt due to Nice-style instabilities are thought to range from ~2-4 km/s (*Morbidelli and Rickman, 2015*). However, to cover impacts that fall outside of this range, and to make our results relevant to populations with slower mutual impact speeds such as the Cold Classical Kuiper Belt (*Greenstreet et al. 2019*), we allow our impact speeds to vary from 0.5 to 5 km/s. Additionally, we assume bodies to have uniform initial temperatures, running suites of simulations at 30K, 50K, and 100K initial temperatures. Assuming that the faint young Sun being only 70% as luminous as the current Sun (*Sagan and Mullen, 1972*) and that each object can be treated as a blackbody in radiative equilibrium with an emissivity of 0.9 (*e.g., Sagan and Mullen, 1972; Stansberry et al. 1996*), these temperatures correspond to heliocentric distances of 76 AU, 27 AU, and 6.9 AU. These distances span the heliocentric distances of the proto-Kuiper Belt, and therefore represent reasonable initial temperatures for our simulations. Nevertheless, to understand the sensitivity of final results to temperature, we also run a suite of simulations at 3km/s, varying the temperature between 30K and 100K in 10K increments.

| Parameter | Range |
| --- | --- |
| Object Temperature | 30K - 100K |
| Object Diameter | 25 - 400 km |
| Impactor/Target Diameter Ratio | 1:1, 1:10, 1:100 |
| Impact Speed | 0.5 - 5 km/s |

**Table 1:** Parameters for Initial Sensitivity Study

We ignore gravity in our simulations, as the surface gravity on a 100 km diameter object with a density of water ice is only ~0.01 m/s$^2$, corresponding to an escape speed of 35 m/s; this is negligible compared to the impact speeds being simulated. Even at the largest sizes considered in our sensitivity study, a 400 km diameter icy body has a surface gravity of only ~0.05 m/s$^2$ and a surface escape speed of ~140 m/s, which is still small compared to the range of impact speeds under consideration. The advantage of neglecting self gravity is a considerable increase in computational speed, allowing us to pursue higher grid resolution (which reduces temperature differences between adjacent cells); this has a larger impact on the amount of AWI that crystallizes due to the exponential dependence of the AWI crystallization timescale on temperature.

We ran each simulation for 40 minutes post-impact, after which the bodies were fully disrupted, and well after parcels of material reached their steady state post impact temperatures (such steady state temperatures are generally reached in the first few minutes). We record thermodynamic properties of the tracers every 0.25 seconds, which is sufficiently fine to resolve the impact shockwave passing through a grid cell. If material leaves the computational mesh, the tracer records a temperature of 0K, halting any subsequent AWI crystallization. This leads to a conservative estimate in the amount of AWI that crystallizes.

3. Results and Discussion

During impacts between amorphous water ice (AWI) rich bodies, we find that, in general, material either remains minimally crystallized (<0.1% AWI crystallizes), or nearly fully crystallizes (>99.9% AWI crystallizes), with very little of the impactor material becoming partially crystallized. This is due to the exponential dependence of the crystallization timescale ($\tau$) on temperature, which varies rapidly as a function of temperature (e.g., decreasing two orders of magnitude from 400 minutes to only 4 minutes between 138 K and 158 K). If a parcel of material is shock heated to lower temperatures (below ~140 K), it will experience minimal crystallization during the impact process; conversely, a parcel heated to higher temperatures (e.g., above 175 K) will fully crystallize in a matter of seconds. There is little material in any of the simulations that is shock heated to an intermediate temperature for an intermediate period of time, such that it only partially crystallizes.

*Sensitivity to Size*

In our simulations between equal-sized objects, we assume that the objects are each 100 km in diameter, consistent with the "born-big" hypothesis (*Morbidelli et al. 2009*). Hydrodynamic scaling requires that the time material spends in a shocked state scales directly with impactor size; we therefore expect larger bodies will lead to greater degree of crystallization. We conducted a suite of simulations to determine how sensitive our results would be to the actual initial sizes of the impacting objects. We simulated 3 km/s impacts between pairs of 25 km, 50 km, 100 km, 200 km, and 400 km diameter AWI-rich objects at an initial temperature of 50 K, scaling the resolution of the computational mesh to keep the number of tracers the same and the run times manageable. These choices of size span a factor of 16 in object diameter, corresponding to variations in the volumes of impactors of a factor of 4096. Nevertheless, we find that the amount of AWI that crystallizes (or remains uncrystallized) is extremely similar in each of these simulations.

Furthermore, we can compute a histogram to understand what fraction of, and to what degree, this initial AWI crystallizes. The histograms show that the degree to which these objects crystallize are extremely similar to one another, with nearly all of the initial AWI either minimally crystallizing or nearly fully crystallizing (see figure 3)[3]. Looking more closely at the amount of AWI that remains uncrystallized (less than 0.1% crystallized) and fully crystallizes (greater than 99.9% crystallized), we find that there is minimal dependence in these quantities on object size (see figure 3). This demonstrates that the crystallization behavior of AWI is highly insensitive the absolute sizes of the colliding bodies; even if the initial primordial icy bodies in the solar system had different sizes than suggested by the "Born Big" hypothesis, the results would not change appreciably, and the results of our study would still apply.

The differences in AWI crystallization between the different cases are small, but our results show that larger impactors nevertheless cause slightly more crystallization. This agrees with what is expected from hydrodynamic similarity, i.e. the material shocked by the largest impactors spends about 16 times longer at high temperature than those material at similar location within the smaller bodies. However, the effect is relatively modest and this tells AWI crystallization is much more sensitive to peak temperatures than the time material spends at high temperature and pressures, at least for the impactor size ranges considered here. For much smaller impactors we would expect a drop in the degree of crystallization (e.g. *Bowling et al. 2020*).

---

[3] We note that all histograms from out simulations have this structure in which the vast majority of material is either in the least or most crystallized bins, and therefore omit them in other cases in favor of more useful plots comparing how the amount of AWI that is fully/minimally crystallized depends on the quantify of interest.

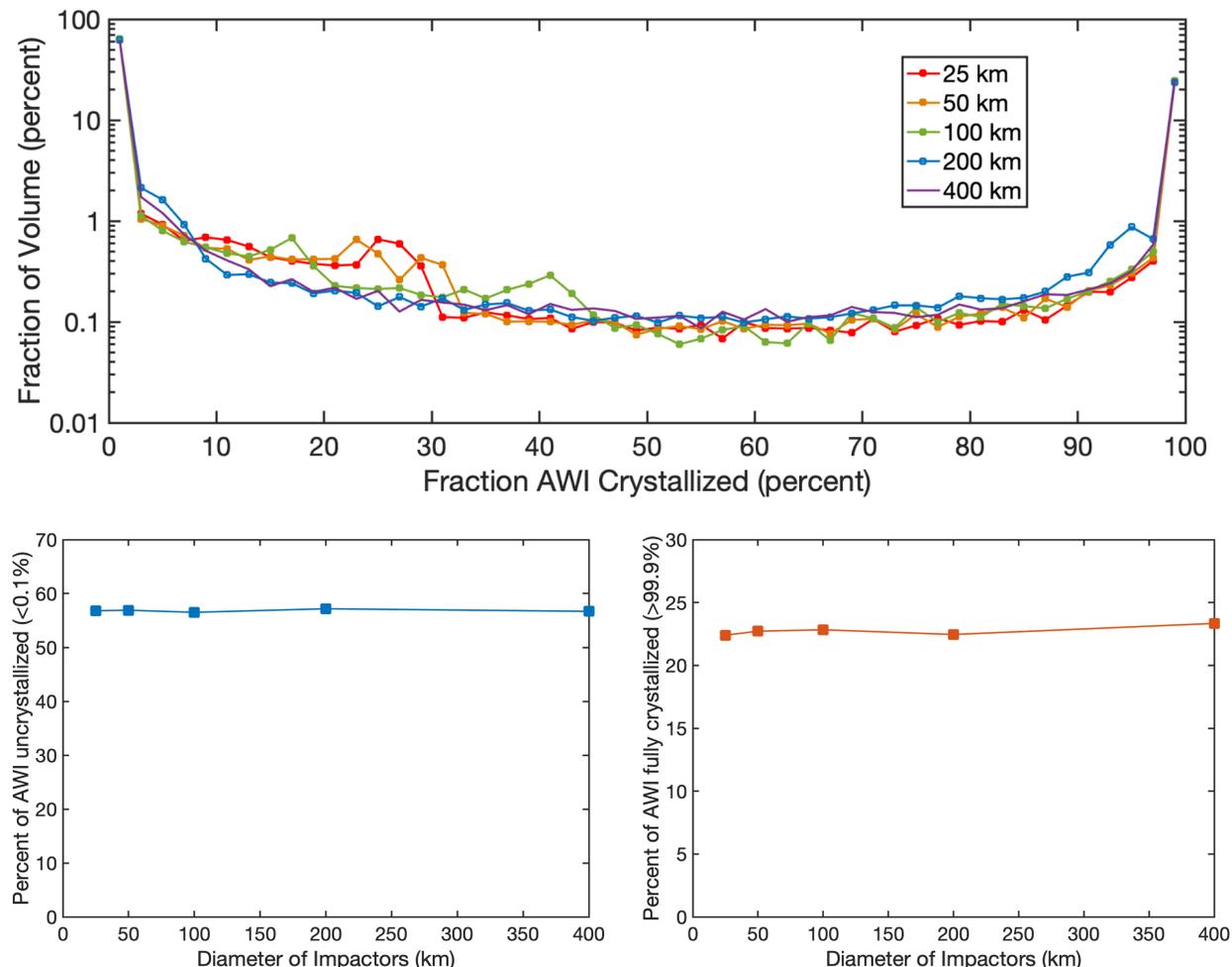

**Figure 3:** *Sensitivity of AWI Crystallization to Object Size.* We simulate two 50 K objects colliding at 3 km/s, with initial diameters of 25, 50, 100, 200, and 400 km. *(Top)* the histograms of AWI crystallization plot the volume fraction of the initial AWI (vertical axis) that is crystallized to each degree, using 50 bins with uniform widths of 2%. Note that the vertical axis is logarithmic. Each histogram shows that minimal material is partially crystallized, and that the vast majority of the initial AWI is either minimally (leftmost bin) or nearly fully crystallized (rightmost bin). The extreme similarities of these histograms revealing that there is little difference in AWI crystallization between these different impactor sizes. We can also plot the volume fraction of initial AWI that remains less than 0.1% crystallized (*bottom left)* or crystallized to at least 99.9% (*bottom right*) as a function of impactor diameter (this is equivalent to using 1000 histogram bins, and looking at only the leftmost and rightmost bins). Both of these plots are nearly horizontal lines, showing that the amount of AWI that survives uncrystallized (to less than 0.1%) or fully crystallizes (to at least 99.9%) is essentially independent of impactor size.

*Sensitivity to Initial Object Temperature*

We explored how the initial temperature of the planetesimal affects the crystallization of amorphous water ice through the impact process. We simulated a series of 100 km impactors colliding at 3 km/s with initial temperatures of 30K, 50K, 60K, 70 K, 80 K, 90K, and 100K. We find that, surprisingly, the initial temperature of the object has little impact on the amount of amorphous water ice that crystallizes. The amount of AWI that fully crystallizes with these different temperatures is similar, lying within a narrow band only ~4 % wide for temperatures of 80 K and below; they differ from the amount of AWI that crystallizes at 60 K by no more than ± ~10%. It is only the higher temperatures (90 K and 100 K) that experience significantly enhanced crystallization, and even then the increase in crystallization is relatively minor. Around 29% of the initial AWI in our 90 K run fully crystallized (1.23 times the crystallization of our 60 K run), and around 35% of the initial AWI in our 100 K run fully crystallized (1.46 times the crystallization of our 60K run; see figure 4). Similarly, the amount of AWI that remains uncrystallized is very weakly dependent upon temperature at 80 K or below and shows a similar ~6% wide range of values, from ~58% at 30K to 52% at 80K; this drops to ~44% at 90 K (only 0.8 times the AWI that survives uncrystallized in our 60 K run) and ~37% at 100 K (only 0.77 times the AWI that survives uncrystallized in our 60 K run; see figure 4).

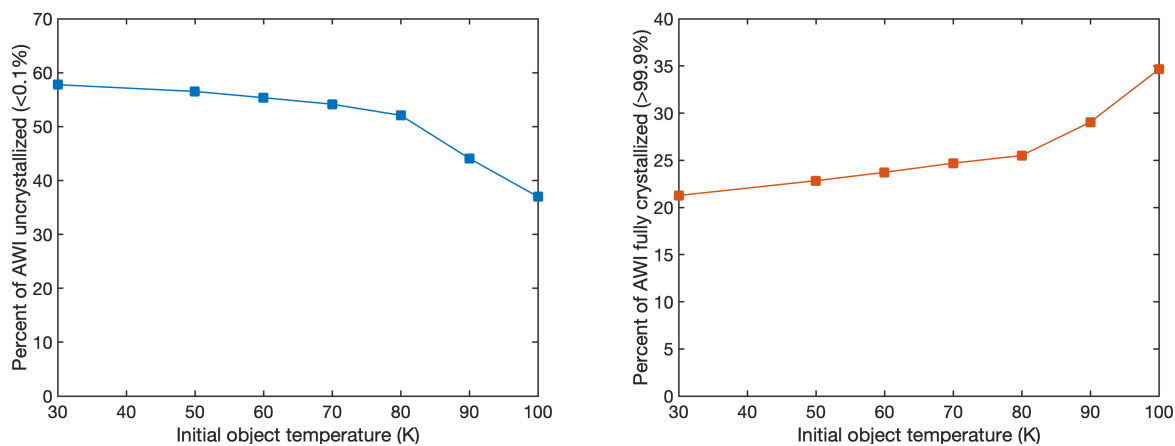

**Figure 4:** *Fraction of initial AWI that remains uncrystallized and fully crystallizes as a function of initial object temperature.* The crystallization dynamics of AWI depends weakly on the initial temperatures of the colliding objects. Above ~80 K, this dependence becomes significant, however such initial temperatures would be characteristic of objects interior to the proto-Edgeworth-Kuiper Belt, suggesting that such initial temperatures are unlikely.

This begs the question: what temperatures are reasonable for proto-Edgeworth-Kuiper Belt (proto-EKB) objects? If their temperatures are in radiative equilibrium with a faint young Sun that is only 70% as luminous as today (*Sagan and Mullen, 1972*), and we assume a blackbody with surface emissivity of 0.9 ( (*e.g., Sagan and Mullen, 1972; Stansberry et al. 1996*), these temperatures of 30 K, 50 K, 60 K, 70 K, 80 K, 90 K, and 100 K correspond to 76, 27, 19, 14, 11, 8.5, and 6.9 AU respectively. This range of heliocentric distances spans the expected range of the proto-Edgeworth-Kuiper Belt, which planet migration scenarios place in the ~20-40 AU range (e.g., *Nesvorný, 2018*); thus these higher temperatures (above 80 K) are unlikely to apply to the proto-EKB. Furthermore, conditions in the protosolar nebula suggest that amorphous water ice only forms in the more distal parts of the Sun's protoplanetary disk, at temperatures below ~90 K (*Ciesla, 2014*)

      Additionally, although the AWI crystallization model of *Kouchi et al. (1994)* shows that the AWI crystallization timescale at 80 K and below is on the order of ~Gyr or longer, the AWI crystallization timescale at 90 K and 100 K are on the order of ~500 kyr and ~1 kyr, respectively. This is much shorter than the timescale for the onset of the giant planet instability and onset of planet migration in Nice-style models (e.g., *Nesvorný, 2018*); thus any material that started at these temperatures would have rapidly crystallized *prior* to experiencing the collisional evolution being studied here. Of note: the timescale for such large bodies to reach radiative equilibrium with the Sun throughout their interiors is much longer than the timescale of the onset of planet migration (*Steckloff et al. 2021; Lisse et al. 2022*), with the interiors of these objects likely cooler than these equilibrium temperatures. Together, these three arguments suggest that, for giant planet instability-driven collisional evolution of the proto-EKB, the amount of AWI that crystallizes is not sensitive to the initial temperature of the object.

    *Sensitivity to Impact Speed*

      We also explore how different impact speeds affect AWI crystallization. We consider relative impact speeds from 500 m/s to 5 km/s, and initial temperatures of 30 K, 50 K, and 100 K, and find that AWI crystallization is highly sensitive to impact speed across this range (see figure 5). At the slow end of this range (1 km/s and slower), negligible AWI crystallizes regardless of initial temperature. At the other extreme, no AWI survives uncrystallized, and nearly all AWI (~90%) fully crystallizes at 5 km/s (see figure 6). Extrapolating, it is clear that impacts slower than 500 m/s will also preserve essentially all AWI, while impacts faster than 5 km/s will cause nearly all AWI to fully crystallize. Non-zero porosity is likely to increase the amount of AWI that crystallizes; however this is a topic for our follow-up work. Regardless, it is clear that the amount of AWI that crystallizes/survives is highly dependent on the impact speed between 1 km/s and 5 km/s.

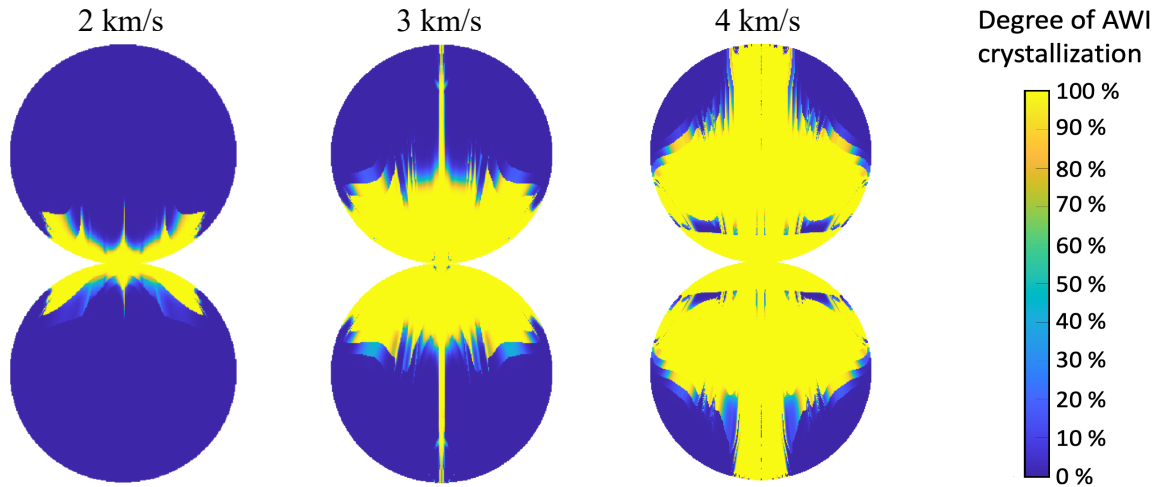

**Figure 5:** *AWI crystallization fraction provenance for two 100km objects striking each other at 50K initial temperature.* The amount of AWI that crystallizes is highly sensitive to impact speed. (*Left*) at 2 km/s, minimal AWI is crystallized, mostly near the point of contact; (*center*) this region of crystallization expands at 3 km/s. *(Right)* at 4 km/s, the objects nearly fully crystallize. The "spike" near the symmetry axis (most visible in the 3 km/s case) is most likely a numerical artifact that results from the extremely small cell volumes near the axis of symmetry. Ultimately, their volumes are very small, and thus do not significantly affect our results.

Crucially, our results show that there is a wide range of outcomes between 2 km/s and 4 km/s, which is the expected variation in impact speeds during the early catastrophic collisional evolution of the proto-EKB following Nice-style instabilities (*Morbidelli and Rickman, 2015*). This suggests that the fragments and reaccreted bodies that result from this collisional evolution, which are thought to comprise the modern day population of comets, centaurs, and TNOs, are likely to exhibit significant stochastic variation in the preservation of any amorphous water ice present. Thus, whether or not primordial AWI survived this collisional evolution depends sensitively on each object's collisional history.

Furthermore, different parts of the proto-EKB have different average impact speeds, resulting in a rate of AWI survival/crystallization that varies with location in the disk. Prior to the Giant Planet instability, the mean collisional velocities between proto-EKB objects varied from 780 m/s interior to 20 AU, down to 240 m/s beyond 25 AU (*Morbidelli and Rickman, 2015*); these impact speeds are too slow to trigger significant AWI crystallization, suggesting that any primordial AWI would likely survive to the Giant Planet instability. At the time of the Giant Planet instability and onset of planet migration, collisional speeds are likely to be higher in the inner parts of the disk relative to the outer parts, likely producing a heliocentric gradient in the survival of AWI; with the amount of AWI surviving any average collision increasing with heliocentric distance.

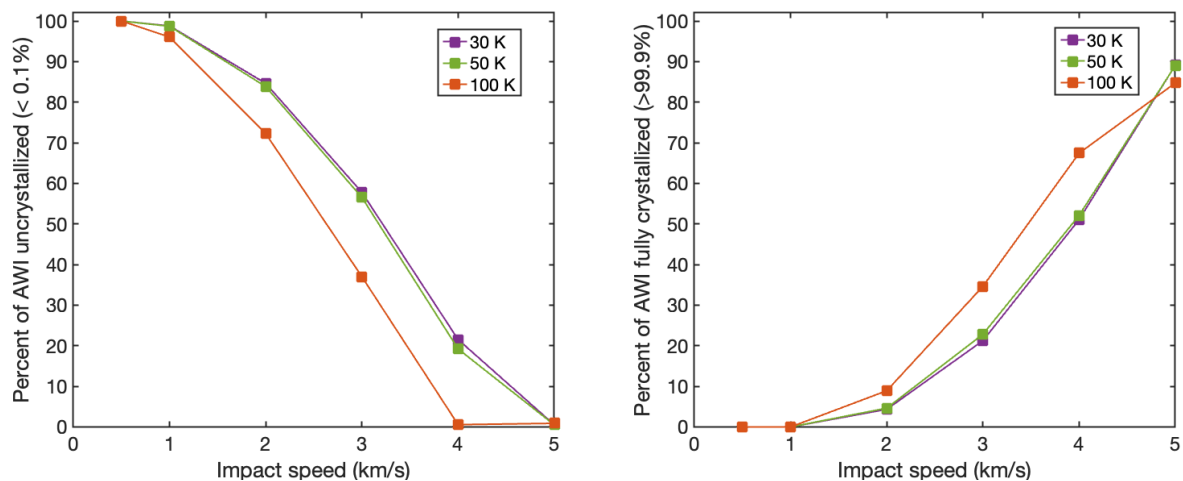

**Figure 6:** *Fraction of initial AWI that remains uncrystallized and fully crystallizes as a function of impact speed for different initial temperatures.* The crystallization dynamics of AWI is highly sensitive to impact speed across all temperatures considered. The amount of AWI that crystallizes is particularly sensitive to the range of expected impact speeds between icy bodies during planet migration of 2-4 km/s (*Morbidelli and Rickman, 2015*).

    The number of energetic collisions experienced by a parcel of AWI can significantly alter the amount of AWI that survives, as subsequent catastrophic impacts will continue to crystallize large fractions of AWI. Objects in the inner part of the disk are expected to experience more disruptive collisions than in outer parts of the disk (*Morbidelli and Rickman, 2015*), which would further reinforce this general trend of increasing AWI survival with increasing heliocentric distance. For the size of body considered here, *Morbidelli and Rickman (2015)* compute that in the inner proto-EKB (< 20 AU), all objects of this size are likely to experience one or more catastrophic impacts; however objects 100 km in diameter in the outer parts of the disk (>25 AU) may avoid such catastrophic impacts altogether. The exact likelihood of AWI survival depends sensitively on the mass density and size frequency distribution (SFD) of the proto-EKB population. Shallower SFDs (lower SFD power law indices) predict more catastrophic collisions, while steeper SFDs (higher SFD power law indices) tend to produce fewer such collisions. Unfortunately, these quantities are poorly constrained. Without a better understanding of the SFD and collisional dynamics of the proto-EKB, it is presently unclear how much AWI survives and is emplaced in the comet reservoir populations (i.e., Oort Cloud and TNO populations). Nevertheless, if AWI does survive, it is more likely to have originated in the outer parts of the proto-EKB.

    Finally, collision velocities are generally too low within the present Oort Cloud (*Stern, 1988*) and TNO (*Durda and Stern, 2000; Abedin et al. 2021*) populations to trigger significant AWI crystallization. Thus, any AWI that survives this early collisional processing is likely to experience no further significant impact-driven AWI crystallization over the age of the Solar

System. Lastly, during their migration through the Centaur region, ecliptic comets experience negligible collisional evolution prior to entering the Jupiter Family Comet population (*Durda and Stern, 2000*). Therefore, if any AWI is present in the bulk nuclei of short period comets, it is likely to be primordial AWI that survived largely unaltered to the present day.

*Impacts Between Unequally Sized Objects*

We also considered impacts between objects of unequal sizes, and simulate a 10 km diameter impactor striking a 100 km diameter target body. As with impacts between equal-sized bodies, the amount of AWI that crystallizes is highly sensitive to impact speed (see figure 7); at lower speeds (2 km/s or less), negligible amounts of AWI crystallize. However as the impact speed increases, more AWI crystallizes, with the impactor nearly fully crystallizing at impact speeds of 4 km/s and higher (see figure 8). Nevertheless, in these cases of unequally sized objects, the amount of target material affected by the impact process is comparable in volume to the impactor itself; even at higher impact speeds that fully crystallize the impactor, the total amount of AWI that crystallizes is approximately double the volume of the impactor alone. In the case of a 10 km object striking an otherwise identical 100 km target, the target has 1000 times the volume of the impactor. As a result, a single such impact would only affect ~0.1% of the AWI present in the target body, even if the impact ultimately disrupts the target itself. This is in agreement with the impact simulations of *Schwartz et al. (2018)*, which found that catastrophic impacts between icy bodies up to ~1 km/s generate negligible heating over the vast majority of the material in the impactor and target.

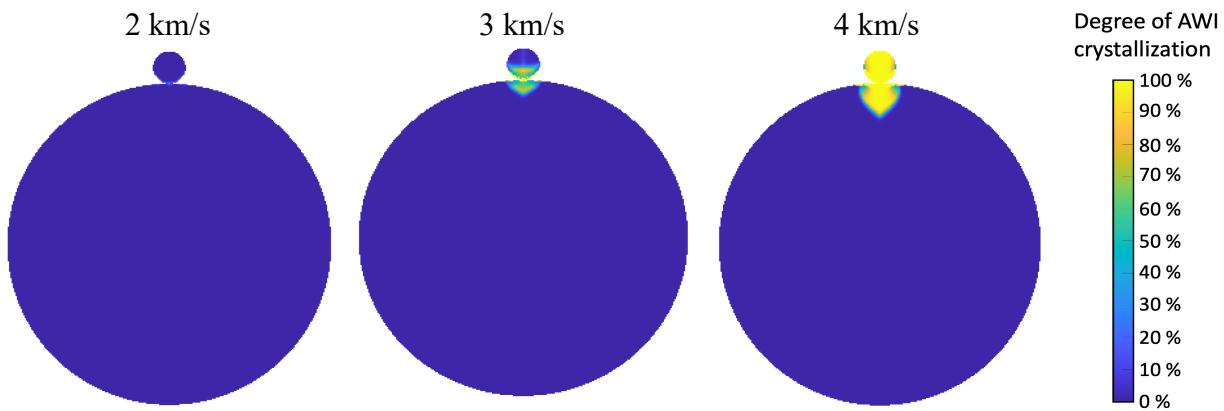

**Figure 7:** *AWI crystallization fraction provenance for 10km impactors striking 100km targets at 50K initial temperature.* The amount of AWI that crystallizes is again highly sensitive to impact speed. (*Left*) at 2 km/s, minimal AWI is crystallized; *(center)* at 3 km/s, the impactor is partially crystallized, along with a comparable volume of the target. *(Right)* at 4 km/s, the impactor is nearly fully crystallized, and fully crystallized a volume of the target comparable in volume to the impactor. In all cases, AWI crystallization is localized to the impact site itself, with negligible AWI crystallization observed distal to the impact site.

The relative frequency with which 10 km sized impactors collide with 100 km target objects depends sensitively on the size-frequency distribution (SFD) of the collisional population. To first order, the SFDs of such populations are well represented by power laws, where the number of objects ($dN(x)$) with diameters between $x$ and $x + dx$ is described by

$$dN(x) = Cx^{-q}dx \qquad (7)$$

where $C$ is a constant[4]. At a power law index of $q = 3$, the SFD is flat, such that equal amounts of material (by volume) are found in each size range between $x$ and $x + dx$; a flat SFD would (to zeroth order) result in approximately ~1000 impactors with diameters of 10 km objects striking 100 km targets for every 100 km impactor. However, because the impact process results in the impactor thermally affecting a volume of material in the target that is approximately equal to its own volume, the cumulative effect of 10 km impactors could be comparable to the cumulative effect of a 100 km impactor; this assumes that these ~1000 impactors with diameters of 10 km keep the target either well gardened, or disrupt it, such that each impactor is able to strike "fresh" (i.e., uncrystallized) material that previous impacts dredged up or otherwise exposed from deep in the target's interior.

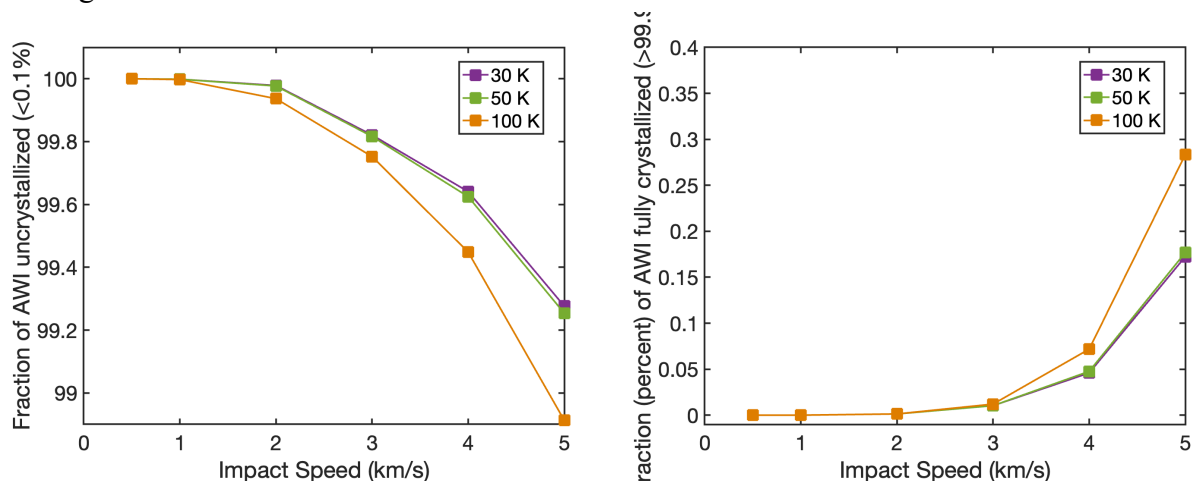

**Figure 8:** *Fraction of initial AWI that remains uncrystallized and fully crystallizes during impact between 100km and 10km objects as a function of impact speed for different initial temperatures.* The crystallization dynamics of AWI is highly sensitive to impact speed across all temperatures considered. The total amount of AWI that crystallizes is much lower than impacts between 100km objects, as the impactor penetrates and crystallizes material to a depth comparable to its own size. Thus, the majority of the target is unaffected. Nevertheless, such smaller impactors are much more common than the larger, catastrophic impacts.

We also simulate small (1 km diameter) impactors, striking a much larger target (treated as a flat surface) with iSALE. Furthermore, these impactors cause negligible AWI crystallization at slower speeds (3km/s and below), but crystallize a volume of AWI comparable to the volume of the impactor at higher speeds (4 km/s and faster). Unlike 10 km impactors, 1 km impactors are too small to disrupt 100 km target objects. As a result, the cumulative effect of these smaller impacts is to crystallize only the surface layers of the target object, down to the impact gardening depth (which is comparable to the size of the impactor itself). Thus, small impactors could leave the interiors of these 100 km objects largely thermally unaltered yet would cumulatively fully crystallize any AWI present in the outer ~km of the object. This is consistent with the results of *Porter et al. (2010)*, who found that micrometeorites in the outer solar system would anneal surface ices of EKB objects, fully crystallizing them.

---

[4] This is the functional form of a *differential* size frequency distribution. Other studies report *cumulative* size frequency distributions, which are the cumulative, integrated number of objects smaller than x. Differential and cumulative SFDs that describe the same populations have power law indices ($q$) that differ by a value of 1, due to their differential/integral relationship with one another.

In the case that AWI rich objects are only impacted by such small impactors, their interiors could hide AWI deposits that could survive to the present day. However, the dynamical modeling of *Morbidelli and Rickman (2015)* suggest that this scenario is very unlikely, and that larger impactors are very likely to disrupt the vast majority of primordial 100 km objects for most reasonable values of the SFD power law index ($q$ = 2-5 - 3.5), and therefore the presence/absence of the AWI in icy bodies is determined by impactors sufficiently large to disrupt the target body into fragments.

We note that, although our simulations are finished after 40 minutes, residual impact heat can remain within the impact fragments or near the impact size for potentially years or centuries (*e.g., Yasui et al. 2021*), potentially causing additional AWI crystallization. This process is likely to affect the materials that only partially crystallize in our simulations (between 0.1% and 99.9% crystallization), as the uncrystallized materials are much too cool for appreciable crystallization and the fully crystallized materials have no further AWI to crystallize. It is unclear presently how much additional crystallization this long-term cooling would produce, which would require additional numerical tools that are outside the scope of this study. Nevertheless, the vast majority of the material across all simulations is either uncrystallized or fully crystallized, with only ~10% of the material being partially crystallized; thus, the uncertainties that would result are at most on this order. Regardless, this process further underscores that our results are conservative on the amount of AWI that crystallizes due to impacts.

Impacts are not the only known process that can affect AWI content; cosmic ray exposure can affect the abundance of AWI in these icy bodies (*Dartois et al. 2015; Maggiolo et al. 2020*). Galactic cosmic rays are highly energetic particles (mostly protons, alpha particles, and electrons/positrons) with typical energies on the order of ~GeV (~$10^{-10}$ J), but cosmic rays with energies of up to 3.2 x $10^{11}$ GeV (~51 J) have been detected (*Birt et al. 1995*). These energies are sufficient to knock water molecules out of position in the crystal lattice, creating amorphous water ice (*Leto and Baratta, 2003*). Moreover, cosmic rays are prevalent enough to fully amorphize crystalline water ice near the surface of icy bodies over timescales of only ~60 Myr (*Maggiolo et al. 2020*). However, cosmic rays are adsorbed as they pass through the material of the icy body, and therefore only affect the outer few ~10s of meters of material (*Maggiolo et al. 2020*), leaving the interior of the icy body unaffected. It is presently unclear if the small impact flux into present EKB and Oort Cloud objects is sufficient in the present solar system to overcome cosmic ray amorphization and crystallize the near surfaces of these objects; such an investigation is beyond the scope of this work and left for future studies. Nevertheless, if AWI is found in the deep interiors of these icy bodies, it is shielded from cosmic rays by the overlying material, and is therefore most likely primordial AWI that survived collisional shock heating.

Additionally, we note that radioactive heating early in the Solar System may induce sufficiently high temperatures within icy bodies to crystallize significant amounts of AWI. In general, the most important radionuclides for radiogenic heating have short half-lives on the order of ~1-2 Myr ($^{26}$Al and $^{60}$Fe) and an initial abundance that, while uncertain, is tied to an object's bulk refractory (non-ice) mass fraction (*Prialnik et al. 2004*). Thus, within a few Myr of formation, radiogenic heating of an icy object ceases, regardless of its size.

The net thermal energy deposition results from a balance between heat loss through the surface by advection and thermal radiation (a rate that scales with surface area) and heat production within the object (a rate that scales with volume). Thus, the thermal evolution via radiogenic heating is highly size-dependent; the smaller the object's radius, the higher the surface area-to-volume ratio and the lower the maximum temperature it can attain. This results in a critical size below which the maximum temperature is insufficient to trigger significant AWI crystallization. Indeed, previous studies have round that radiogenic thermal alteration of icy bodies is minor, even negligible, for objects smaller than 10-15 km in radius, especially for compositions rich in water ice and mixed crystalline and amorphous form (e.g., *De Sanctis et al. 2001; Choi et al. 2002; Prialnik et al. 2008b, Sarid & Prialnik 2009*). Thus, the smaller impactors considered here (1 km and 10 km in radius) are too small for radiogenic alteration of their primordial AWI.

For larger objects such as the 100 km objects considered in this work, heat from radioactive decay is more efficiently retained toward their centers. Thus, these larger bodies may have already lost large amounts of the primordial AWI contained in their interiors. In contrast, efficient cooling closer to the surfaces of icy bodies leads to significantly less thermal alteration (and thus AWI crystallization) in their surface and near-surface regions. These unaltered surface and near-surface regions are most susceptible to impact processing, as we show in our current study; as a result, smaller impactors would preferentially crystallize the AWI in the target object that is least susceptible to crystallization via radiogenic heating. Ultimately, the effects of radiogenic heating on AWI crystallization, while still relatively poorly understood, generally work to further enhance the loss of AWI in the early solar system, rendering it more difficult for primordial AWI to survive to the present day.

Finally, disruption events of solar system comets may offer an opportunity to examine the deep interiors of comet nuclei for evidence of the presence/absence of AWI in their deep interiors. For example, the tidal disruption of comet Shoemaker-Levy 9 (*Asphaug & Benz, 1996*) and breakup of Comet 73P/Schwassmann-Wachmann 3 (*Reach et al. 2009*) each produced a series of fragments, many of which undoubtedly came from the comet's deep interior. If a disrupted comet is sufficiently large, it may retain significant AWI deposits in its deep interior (*Lisse et al. 2022*), suggesting that if fragments are observed immediately after such a disruption event, interior fragments will have little time to further evolve, providing an opportunity to directly look for signatures of AWI in the interiors of nuclei. Ideal targets for such an investigation would likely be objects in the dynamical "Gateway" near Jupiter (*Sarid et al. 2019; Steckloff et al. 2020; Seligman et al. 2021*) as such orbits facilitate the rapid migration of Centaurs into the inner solar system, where sublimation torques can more rapidly drive comet splitting and disruption (*Steckloff and Jacobson, 2016; Jewitt et al. 2020*). The probability of detecting AWI in the interior of a freshly disrupted comet is nevertheless quite low, as the nucleus likely needs to be rather large for AWI to even survive to the present day against the slow crystallization induced by solar heating (*Lisse et al. 2022*), which would make the nucleus relatively unsusceptible to torques capable of spinning the nucleus up to disruption.

4. **Conclusions**

We use the iSALE hydrocode to simulate impacts between equally sized icy bodies in the proto-Edgeworth-Kuiper Belt (proto-EKB) and modern day reservoir populations, and a script to compute the degree and quantity of initial amorphous water ice (AWI) that would crystallize during these impacts. We find that the amount of AWI that crystallizes/survives is independent of the absolute sizes of the objects, and weakly sensitive to their initial temperatures. Furthermore, the warmer temperatures where AWI crystallization is sensitive to initial temperature are themselves too warm for AWI to have initially formed and/or survived to the onset of planet migration.

We find that the amount of AWI that crystallizes is highly sensitive to impact speed. In the modern EKB and Oort Cloud, typical impact speeds are sufficiently low that negligible AWI crystallizes. Impact speeds between proto-EKB objects *prior* to the Giant Planet instability and onset of planet migration were similarly too low to enable significant AWI crystallization. However, during planet migration, typical impact speeds of ~2-4 km/s were sufficient to enable significant crystallization of AWI. The amount and degree of crystallization is highly sensitive to these expected impact speeds, with slower speeds of ~2 km/s allowing most initial AWI to survive; higher typical speeds of ~4 km/s result in the majority of AWI fully crystallizing. Ultimately, subsequent impact events during planet migration could cause further crystallization of the surviving AWI; the frequency of such impacts depends on the mass density and size frequency distribution of the proto-EKB, which is poorly understood.

We find similar results for impacts between differently sized objects. Whereas higher impact speeds ~4 km/s fully crystallize AWI in the impactor and a comparable volume of material in the target body, slower impact speeds enable the majority of the AWI in the impactor and impact site to survive. Although a 10 km impactor can significantly deform and partially disrupt a 100 km target body in these impact speed ranges, a 1 km impactor is much too small to cause such disruption. These smaller impactors would therefore only be able to crystallize AWI down to the impact gardening depth. Nevertheless, it is unlikely that such primordial bodies would escape all impacts capable of causing catastrophic disruption. Ultimately, the survival of primordial AWI during planet migration is a stochastic process that is highly sensitive to the impact history of the body; these impact histories require further study to understand how much primordial AWI could survive to the present day.

5.  **Acknowledgements**

We gratefully acknowledge the developers of iSALE-2D, including Gareth Collins, Kai Wünnemann, Dirk Elbeshausen, Tom Davison, Boris Ivanov, and Jay Melosh. Some plots in this work were created with the pySALEPlot tool written by Tom Davison. J.K.S. and G.S. acknowledge support from NASA award 80NSSC18K0497.

Early Collisional Evolution on Amorphous Ice                                    Steckloff et al.Pierazzo, E., N. Artemieva, E. Asphaug, E. C. Baldwin, J. Cazamias, R. Coker *et al.* (2008) Validation of numerical codes for impact and explosion cratering: Impacts on strengthless and metal targets. *Meteoritics and Planetary Science*, 43, 1917–1938.

Porter, S.B.; Desch, S.J.; Cook. J.C. (2010) Micrometeorite impact annealing of ice in the outer Solar System. *Icarus* 208, 492 - 498

Prialnik, D., M. F. A'Hearn, & K. J. Meech, (2008a) A mechanism for short-lived cometary outbursts at sunrise as observed by Deep Impact on 9P/Tempel 1. *Monthly Notices of the Royal Astronomical Society*, 388, L20–L23.

Prialnik, D. & A. Bar-Nun, (1992) Crystallization of amorphous ice as the cause of Comet P/Halley's outburst at 14 AU. *Astronomy and Astrophysics*, 258, L9–L12.

Prialnik, D., J. Benkhoff, & M. Podolak (2004) *Modeling the structure and activity of comet nuclei*, University of Arizona Press, Tucson. pp. 359–387.

Prialnik, D., G. Sarid, E. D. Rosenberg, & R. Merk (2008b) Thermal and Chemical Evolution of Comet Nuclei and Kuiper Belt Objects. *Space Science Reviews*, 138, 147–164.

Reach, W.T.; Vaubaillon, J.; Kelley, M.S.; Lisse, C.M.; Sykes, M.V. (2009) Distribution and properties of fragments and debris from the split Comet 73P/Schwassmann-Wachmann 3 as revealed by *Spitzer* Space Telescope. *Icarus* 203, 571 – 588

Robbins, S.J.; Singer, K.N. (2021) Pluto and Charon Impact Crater Populations: Reconciling Different Results. *PSJ* 2:192 (9pp)

Safrit, T.K.; Steckloff, J.K.; Bosh, A.S.; Nesvorny, D.; Walsh, K.; Brasser, R.; Minton, D.A. (2021) The Formation of Bilobate Comet Shapes through Sublimative Torques. *Planetary Science Journal* 2:14 (10pp)

Sagan, C.; Mullen, G. (1972) Earth and Mars: Evolution of Atmospheres and Surface Temperatures. *Science,* 177, 52 - 56

Sarid, G., D. Prialnik, K. J. Meech, J. Pittichová, & T. L. Farnham (2005) Thermal Evolution and Activity of Comet 9P/Tempel 1 and Simulation of a Deep Impact. *Publications of the Astronomical Society of the Pacific*, 117, 796–809.

Sarid, G., Prialnik, D. (2009) From KBOs to Centaurs: The thermal connection. Meteoritics and Planetary Science, 44, 1905–1916.28

Early Collisional Evolution on Amorphous Ice                                     Steckloff et al.Sarid, G.; Volk, K.; Steckloff, J.K.; Harris, W.; Womack, M.; Woodney, L.M. (2019) 29P/Schwassmann-Wachmann 1, a centaur in the gateway to the Jupiter-family comets. *ApJ Letters* 883, L25 (7 pp.)

Schwartz, S.R.; Michel, P.; Jutzi, M.; Marchi, S.; Zhang, Y.; Richardson, D.C. (2018) Catastrophic disruptions as the origin of bilobate comets. *Nature Astronomy* 2, 379 - 382

Sekanina, Z. and Yeomans, D.K. (1985) Orbital motion, nucleus precession and splitting of periodic Comet Brooks 2. *AJ* 90, 2335 - 2352

Sekanina, Z. (2009) Crystallization of Gas-Laden Amorphous Water Ice, Activated by Heat Transport to its Subsurface Reservoirs, as Trigger of Huge Explosions of Comet 17P/Holmes. *International Comet Quarterly*, 31, 99–124.

Seligman, D.Z.; Kratter, K.M.; Levine, W.G.; Jedicke, R. (2021) A Sublime Opportunity: The Dynamics of Transitioning Cometary Bodies and the Feasibility of In Situ Observations of the Evolution of Their Activity. *PSJ* 2:234 (18 pp.)

Senft, L.E.; Stewart, S.T. (2011) Modeling the morphological diversity of impact craters on icy satellites. *Icarus* 214, 67 – 81

Singer, K.N.; McKinnon, W.B.; Gladman, B.; Greenstreet, S.; Bierhaus, E.B.; Stern, S.A. *et al.* (2019) Impact craters on Pluto and Charon indicate a deficit of small Kuiper belt objects. *Science* 363, 955 - 959

Smoluchowski, R. (1981) Amorphous ice and the behavior of cometary nuclei. *Astrophysical Journal Letters*, 244, L31–L34.

Speedy, R. J., P. G. Debenedetti, R. S. Smith, C. Huang, & B. D. Kay (1996) The evaporation rate, free energy, and entropy of amorphous water at 150 K. *Journal of Chemical Physics*, 105, 240–244.

Stansberry, J.A.; Pisano, D.J.; Yelle, R.V. (1996) The emissivity of volatile ices on Triton and Pluto. *Planetary and Space Science* 44, 945 - 955

Steckloff, J.K.; Johnson, B.C.; Bowling, T.; Melosh, H.J.; Minton, D.; Lisse, C.M.; Battams, K. (2015) Dynamic sublimation pressure and the catastrophic breakup of Comet ISON. *Icarus* 258, 430 - 437

Steckloff, J. & H. J. Melosh, (2016) Are Comet Outbursts the Result of Avalanches? In *AAS/Division for Planetary Sciences Meeting Abstracts*, vol. 48 of *AAS/Division for Planetary Sciences Meeting Abstracts*, p. 206.06.
29